\documentstyle[preprint,prl,aps,epsfig]{revtex}     %preprint version
\begin{document}
\draft 

\title{Radiative Muon Capture on Hydrogen and the Induced Pseudoscalar
Coupling}

\author{G.~Jonkmans,$^{4}$\cite{gjaddress} S.~Ahmad,$^{1}$\cite{saaddress}
D.~S.~Armstrong,$^{5}$ G.~Azuelos,$^{3,4}$ W.~Bertl,$^{7}$
M.~Blecher,$^{2}$ C.~Q.~Chen,$^{1}$  P.~Depommier,$^{4}$
B.~C.~Doyle,$^{1,6}$ T.~von~Egidy,$^{3}$\cite{tveaddress}
T.~P.~Gorringe,$^{6}$ P.~Gumplinger,$^{1}$ M.~D.~Hasinoff,$^{1}$
D.~Healey,$^{3}$ A.~J.~Larabee,$^{1}$\cite{ajladdress}
J.~A.~Macdonald,$^{3}$ S.~C.~McDonald,$^{8}$ M.~Munro,$^{8}$
J.-M.~Poutissou,$^{3}$ R.~Poutissou,$^{3}$  B.~C.~Robertson,$^{9}$ 
D.~G.~Sample,$^{1}$ E.~Saettler,$^{1}$ C.~M.~Sigler,$^{2}$
G.~N.~Taylor,$^{8}$ D.~H.~Wright,$^{1,3}$ and N.~S.~Zhang,$^{1}$}

\address{                                                      
 (1) University of British Columbia, Vancouver, B. C., Canada V6T 1Z1 \\
 (2) Virginia Polytechnic Institute and State University, Blacksburg, VA
24061 \\
 (3) TRIUMF, 4004 Wesbrook Mall, Vancouver, British Columbia, Canada 
V6T 2A3 \\
 (4) Universit\'e de Montr\'eal, Montr\'eal, P.Q., Canada H3C 3J7 \\
 (5) College of William and Mary, Williamsburg, VA 23187 \\
 (6) University of Kentucky, Lexington, KY 40506 \\ 
 (7) Paul Scherrer Institut, CH-5232 Villigen, Switzerland \\
 (8) University of Melbourne, Parkville, Victoria, Australia 3052 \\
 (9) Queen's University, Kingston, Ontario, Canada K7L 3N6 \\
}

\date{\today}
\maketitle

\begin{abstract}
The first measurement of the elementary process $\mu^{-} p \rightarrow
\nu_ {\mu} n \gamma$ is reported. A photon pair spectrometer was used
to measure the partial branching ratio ($2.10 \pm 0.22) \times
10^{-8}$ for photons of $k > 60$ MeV.  The value of the weak
pseudoscalar coupling constant determined from the partial branching
ratio is $g_{p}(q^{2}=-0.88m_{\mu}^{2}) = (9.8 \pm 0.7 \pm 0.3)
\cdot g_{a}(0)$, where the first error is the quadrature sum of
statistical and systematic uncertainties and the second error is due
to the uncertainty in $\lambda_{op}$, the decay rate of the ortho to
para $p \mu p$ molecule. This value of $g_p$ is $\sim$1.5 times the
prediction of PCAC and pion-pole dominance.
\end{abstract}

\pacs{23.40.-s, 13.10.+q, 23.40.Bw, 11.40.Ha }

In semi-leptonic weak interactions the strong force can, in general,
induce four couplings in addition to the usual vector ($g_{v}$) and
axial vector ($g_{a}$) couplings: weak magnetism ($g_{m}$),
pseudoscalar ($g_{p}$), scalar ($g_{s}$), and tensor ($g_{t}$)
\cite{Grenacs}.  G-parity invariance predicts null values for $g_{s}$
and $g_{t}$, in agreement with experiment
\cite{Grenacs,Morita,Baringer}. The CVC prediction for
$g_{m}$\cite{Gell-mann} also agrees with experiment
\cite{Lee}. Theoretical calculations \cite{Bernard} of 3\% precision
are available for $g_{p}$, but no measurement approaching such
accuracy has been reported.

It is difficult to measure $g_{p}$ in beta decay because of the small
value of the 4-momentum transfer, $q$. In muon capture the
pseudoscalar coupling contribution is larger.  Capture on the proton
allows a reliable calculation since this case is free of nuclear
structure uncertainties. Assuming partial conservation of the axial
current (PCAC) and pion-pole dominance of the pseudoscalar coupling,
$g_{p}$ can be related to $g_{a}(0)$ \cite{Goldberger}.  In Fearing's
notation \cite{Fearing},
\begin{eqnarray}
g_{p}(q^{2}) & = & 6.77 g_{a}(0)
\frac{(m_{\pi}^{2} + 0.88 m_{\mu}^{2})}{(m_{\pi}^{2} - q^{2})},
\end{eqnarray}
where $m_{\pi}$ is the charged-pion mass and $m_{\mu}$ is the muon
mass.  In ordinary muon capture (OMC), $\mu^{-} p \rightarrow
\nu_{\mu} n$, $q^{2} = -0.88 m_{\mu}^{2}$, which is far from the pion
pole. Thus $g_a$ dominates $g_p$ and a rate measurement of 4\%
precision \cite{Bardin} yields a 40\% uncertainty in
$g_{p}$. Combining many different OMC results yields only 25\%
precision \cite{Bardin} for $g_{p}$.

In radiative muon capture (RMC), $\mu^{-} p \rightarrow \nu_{\mu} n
\gamma$, $q^{2}$ can be much closer to the pion pole, since $q^{2} =
m_{\mu}^{2}$ at the maximum photon energy ($k \sim 100$ MeV).  The
$g_{p}$ contribution to the RMC amplitude is then more than 3 times
its contribution to the OMC amplitude.  Offsetting this long known,
promising sensitivity to $g_{p}$ is the fact that, in hydrogen, the
partial branching ratio is $\sim 10^{-8}$.  Due to this small
probability and the many potentially large background sources, no
previous measurement has been attempted.

The experiment was performed using the M9A muon beam at the TRIUMF
cyclotron.  Negative muons at 63 MeV/$c$ were selected by an rf
separator, counted by four beam scintillators, and stopped in a liquid
hydrogen target. The $\mu^-$ stop rate was $\sim$6.5$\times$10$^5 {\rm
s}^{-1}$ and the $\pi^-$ contamination was $\pi$/$\mu$ $\leq$
2$\times$10$^{-4}$.  Photons emerging from the target were detected by
$\gamma$ $\rightarrow$ $e^+ e^-$ conversion in a cylindrical lead
sheet and analyzed by tracking the $e^+$ and $e^-$ paths in
cylindrical drift and wire chambers.  An axial magnetic field allowed
the determination of the photon energy from the curvature of the $e^+$
and $e^-$ paths.  Concentric segmented cylinders of plastic
scintillators just inside the lead converter radius (the $A$, $A'$,
and $B$ rings), just outside the lead converter radius (the $C$ ring),
and just outside the drift chamber outer radius (the $D$ ring)
provided the photon trigger via the logic $\overline{A} \cdot
\overline{A'} \cdot \overline{B} \cdot C \cdot D$.  Lastly, two layers
of plastic scintillators and drift chambers surrounding the magnet
were used to identify cosmic-ray backgrounds. For more details on the
experimental setup, see Ref. \cite{Wright}.

Even a small $\pi^-$ contamination in the beam could give rise to a
potentially dangerous background, since each $\pi^-$ produces 
photons from the charge exchange or capture reactions:
\begin{eqnarray}
\pi^{-} + p & \rightarrow & \pi^{0} + n \;\; (60.7\%), \\
\nonumber	    &             & \hookrightarrow \gamma + \gamma  \;\;
				       (k \sim 55-83 \; {\rm MeV}), 
\\ \pi^{-} + p & \rightarrow & \gamma + n \;\; (39.3\%, k \sim 129 \; 
{\rm MeV}).
\end{eqnarray}
Although these pion-induced photons were $\sim 10^{4}$ times more
plentiful than RMC photons, they were prompt and could be vetoed by a
timing cut.  They were, however, useful for measurement of the
acceptance and quantitative comparison to Monte Carlo modeling of the
detector. For this measurement the beam was periodically tuned for 81
MeV/$c$ $\pi^{-}$, which stopped at the same place in the target as
the 63 MeV/$c$ $\mu^{-}$. The pion-induced photons were subjected to
the same geometry cuts as were the photons from $\mu^- p$, but not to
the timing cuts.  Excellent agreement with Monte Carlo was obtained
\cite{Wright}, especially in the $\pi^{0}$ region which overlaps the
RMC region, providing an energy calibration to $\sim 0.25$ MeV.

The spectrometer energy resolution, $\sim 11 \%$ FWHM for photons, was
sufficient to eliminate most of the background for $k>60$ MeV from
$\mu^{-}$ decay with internal or external bremsstrahlung, $\mu^{-}
\rightarrow e^{-} \overline{\nu_{e}} \nu_{\mu} \gamma$.  These photons
have a maximum energy of $m_{\mu}$/2, which prevents study of RMC at
lower energies.  Furthermore, $\mu \rightarrow e \overline{\nu} \nu
\gamma$ events could leak from the region $k <$ 60 MeV into the region
$k >$ 60 MeV due to the high-energy tail of the spectrometer
resolution function.  To determine this high-energy tail background
the photon spectrum from $\mu^+$ stops, where decay is present but
capture is absent, was measured. Comparison of the normalized $\mu^+$
and $\mu^-$ photon spectra, accounting for in-flight annihilation
which contributes to the former but not the latter, yielded a (17.2
$\pm$ 2.5)\% high-energy tail background in the RMC spectrum.

Some fraction of the cosmic-ray induced events were not identified in
the cosmic-ray detectors surrounding the spectrometer magnet, and thus
survived the software veto. The residual background was measured under
experimental conditions when the beam was off.  This and any proton
``beam related'' background was measured when the secondary beam was
diverted to other channels.  The combined background photon energy
spectrum, normalized by time and integrated proton current, agrees
with the spectrum, $100 \! < \! k \! < \! 200$ MeV, produced by the
$\mu^{-}$ beam.  This energy range is above the RMC range and allowed
a reliable extrapolation and subtraction of these backgrounds, $(10
\pm 4)\%$, in the RMC region.

Muon capture occurs on the Au flask containing the liquid H and on the
surrounding Ag heat shields much more rapidly than on H. Au was chosen
for the flask material because it can be made into a very pure
thin-walled container. Also, the disappearance time for $\mu^{-}$ in
Au is 73 ns, in Ag 89 ns, while in H it is 2195 ns \cite{suzuki}.
Events which occurred within 365 ns (5$\tau_{Au}^{\mu^-}$) following
the last $\mu^{-}$ stop were ``blanked'' in software. A fit to the
unblanked time spectrum for $k>60$ MeV enabled a determination of the
part of this background which occured more than 365 ns after a stop.
The net background from the Au and Ag was $(10 \pm 4)\%$.

Impurities of deuterium or heavy elements in the liquid hydrogen could
have been a serious source of backgrounds.  In liquid the $\mu^{-}$ is
rapidly captured into the singlet atomic state. This small neutral
system quickly forms a $p \mu p$ molecule, usually in the ortho state,
which can decay ($\lambda_{op} = (4.1 \pm 1.4) \times 10^{4} {\rm
s}^{-1}$ \cite{Bardin}) to the para state. However, if there are
contaminants the $\mu^{-}$ can transfer from the proton to form a
$\mu$A atom. Heavy impurities were reduced to $\ll 10^{-9}$ by
thorough bakeout and pumping, and by passing the gas through a
palladium filter \cite{Bertl}. If deuterium is present a $\mu d$ atom
can be formed, followed by a $p \mu d$ molecule.  The latter can form
a $\mu$$^{3}$He system via muon-induced fusion.  Natural H contains
$>$ 100 ppm deuterium, and since capture on $^{3}$He occurs at a much
higher rate than on H, the background from $^{3}$He RMC would be
larger than the H signal.  Our liquid was obtained from the
electrolysis of two samples of deuterium-depleted H$_{2}$O.  For half
of the data-taking it contained $1.4 \pm 0.2$ ppm deuterium
\cite{Metabolic}, while for the other half this contamination was $ <
0.1$ ppm. A $(1.1 \pm 0.3)\%$ $^{3}$He RMC background was 
determined from a short run with natural H.

Other background sources were explored. These include: 1) $\pi^-$ and
$\mu^-$ capture in the Pb beam collimators without hits in the beam
counters. As these collimators were located quite far from the target
such events were vetoed by geometry cuts. 2) Beam $\pi^-$ could decay
in flight to high-energy $\mu^-$ which would capture on the Au/Cu
backplate of the target. The beam simulation which reproduced the
observed stopping distribution indicated that the number of such
events was small and a geometry cut further reduced them.  3) $\mu^-$
which stopped in the veto scintillators would capture on C which has a
disappearance time similar to H. The stopping simulation indicated
this background was $< 1\%$. 4) Before forming a $p \mu p$ molecule
the $\mu p$ atom could diffuse into the Au and capture after a long
time, thereby defeating the blank cut.  The beam simulation showed
very few stops within 2 mm of the Au flask.  The mean free path of
such atoms is $\ll$ 1 mm and thus this background is negligible.  5)
Multiple single tracks in near-coincidence could imitate high-energy
pairs.  However, all trigger counters were connected to multihit TDCs,
which allowed a straightforward veto of these spurious events.  The
net contribution from all these background sources was $(3 \pm 2)$\%.

Data taking took place over a 3.5 year period. The total number of
$\mu^-$ stops examined was $3.6 \times 10^{12}$.  The uncut photon
energy spectrum is shown in Fig.\ 1. The peaks are from bremsstrahlung
and pion-induced reactions. The shaded region shows the spectrum after
application of the prompt cut, indicating elimination of pion
events. The number of RMC photons observed ($k>60$ MeV), after all
cuts and background subtractions, is $N_{0}=279 \pm$26.

The open histogram in Fig.~2 shows the photon spectrum after geometry,
timing, cosmic ray, and energy ($k\! > \! 60$ MeV) cuts were
imposed. The overlayed dark shaded region shows the remaining
bremsstrahlung background, obtained from the normalized $\mu^{+}$
spectrum (with annihilation-in-flight removed).  The light shaded
region shows the photon spectrum after the residual Au, Ag, $^{3}$He,
cosmic ray, ``beam-related'' and bremsstrahlung backgrounds have been
removed.  This represents the first observation of a signal from RMC
on hydrogen.

In order to extract $g_{p}$ from the data, a calculation of the RMC
branching ratio $R_{\gamma}$ and photon energy spectrum in terms of
$g_{p}$ was made using the theory of Fearing and Beder
\cite{Beder}. This is a relativistic approach based on tree-level
Feynman diagrams. Monte Carlo events were generated from the
theoretical photon spectrum for $k\!  >\! 50$ MeV, $t>$365 ns, and
several values of $g_{p}$, and then analyzed like experimental events.
The $q^2$ dependence of $g_p$ assumed was that of Eq.~1.  The values
of the partial branching ratio and $g_{p}$ were obtained when the
number of Monte Carlo RMC events $N$ was equal to the number of
experimental RMC events $N_{0}$.

	The number of Monte Carlo events generated was
\begin{eqnarray}
N&=&R_{\gamma}  \kappa
N_{\mu}  \epsilon_{cr}  P  F_{a}.
\end{eqnarray}
$N_{\mu}$ is the number of beam counter coincidences (stops),
corrected for multiple $\mu^{-}$ in a beam burst.  $\epsilon_{cr}$ is
the efficiency of cuts to eliminate multiple single particles and
cosmics (these cuts were not applied to pion-induced photons from
which the acceptance was obtained, nor to simulated photons). It is
$0.93 \pm 0.01$ and was obtained by noting the effect of these cuts on
$\mu^{-}$ beam events which were rejected by the prompt cut,
i.e. $\pi^-$-induced events which came in the $\mu^-$ beam.  Pileup
muons blank a fraction $1-P$ of real and background events which are
captured more than 365 ns after a stop, where $ P = 0.750 \pm 0.011$
was calculated from knowledge of the timing logic and the incident
beam rate. This calculation reproduced the ratio of events before and
after the blank cut, after all other cuts had been imposed.  $F_{a}$
is the ratio of the measured acceptance in each run period to the
simulated acceptance.  The acceptance was independent of beam rate and
within a running period was constant to within a few percent, but
there were variations of up to 10\% between running periods.

The factor $\kappa = 0.88 \pm 0.01$ is a run-independent product of
several small, measured or calculated corrections.  The largest,
$0.926 \pm 0.005$, is the fraction of muons in $N_{\mu}$ that stopped
in the liquid.  It was measured by replacing the liquid with a
scintillator of appropriate shape and counting coincidences with the
beam counters; the measured value agreed with Monte Carlo predictions
of the beam.  Other factors in $\kappa$, each of which had less than a
2\% effect, included small measured inefficiencies in the beam and
trigger counters, random vetoing in the trigger due to the singles
rates in the $A, A', B$ veto scintillators, and muon miscounting
related to electrons in the beam or from the target.

Due to the spin-dependence of the weak interaction, the extraction of
$g_p$ from the data depends on knowledge of the relative fraction of
muons in the ortho and para $p\mu p$ states. Thus the Monte Carlo,
which includes all the muonic atomic and molecular processes, was run
for various values of $\lambda_{op}$. The sensitivity to the rates of
the other atomic and molecular processes was negligible.

Our measured partial branching ratio for photons of $k > 60$ MeV is
$(2.10 \pm 0.20 \pm 0.09) \times 10^{-8}$, where the first error is
statistical and the second systematic.  This branching ratio arises
from the mixture of muonic states relevant in the experimental time
window ($t > 365$ ns): 6.1\% in the singlet $\mu p$ state, 85.4\% in
the ortho $p\mu p$ state and 8.5\% in the para $p\mu p$ state.  The
shape of the energy spectrum in the experimentally-accessible region
is in good agreement with the theoretical prediction (see Fig.\ 2).
The results for $g_p$ are shown in Fig.\ 3, where $g_{p}$/$g_{a}$ is
plotted versus $\lambda_{op}$. The expected theoretical result (Eq.~1)
is $6.77 \cdot g_{a}(0)$, at $q^{2} = -0.88 m_{\mu}^{2}$.  RMC yields
$g_{p}/g_{a}(0) = 9.8 \pm 0.7 \pm 0.3$, a factor $\sim$1.5 times the
expected value, if the experimental value $\lambda_{op}= (4.1 \pm 1.4)
\times 10^{4} {\rm s}^{-1}$ \cite{Bardin} is used.  The first error
includes statistical and systematic errors added in quadrature, while
the second is due to the error in $\lambda_{op}$.  The present result
for $g_p$ is independent of energy cut, $k > 60$ MeV, blank time cut,
$t_{b} > 365$ ns, and is not very sensitive to $\lambda_{op}$.

In Fig.\ 3, ``Saclay'' refers to the most recent and accurate OMC
measurement \cite{Bardin}, while ``World'' refers to the world average
of OMC (including older bubble chamber) data \cite{Bardin}.  It should
be noted that this average was obtained from many different
experiments with widely varying conditions and results.  Within large
errors, OMC is in agreement with the expected value of $g_{p}$ if one
adopts the measured value of $\lambda_{op}$.  However, the OMC results
are generally more sensitive to $\lambda_{op}$ than the present
measurement, especially the ``Saclay'' result, and the extracted value
of $g_p$ would decrease significantly if the theoretical value
$\lambda_{op} = (7.1 \pm 1.2)\times 10^{4} {\rm s}^{-1}$
\cite{Bakalov} is correct.

In conclusion, we have made the first measurement of the elementary
radiative muon capture process, and determined a partial branching
ratio of $(2.10 \pm 0.22) \times 10^{-8}$, for photons of $k > 60$
MeV.  We have assumed that the measured values of $\lambda_{op}$ and
other $\mu$H parameters are correct, the $q^2$ dependence of $g_{p}$
is given by (1), the $R_{\gamma}$ and energy spectrum for RMC on H are
correctly predicted by a relativistic perturbation theory calculation
\cite{Beder} based on tree-level Feynman diagrams, and that a large
background has not eluded detection in this experiment.  Based on
these assumptions, a substantial deviation of $g_{p}$ from the
predicted value is obtained.

% Acknowledgements: 
We wish to thank the TRIUMF hydrogen target group, in particular
A. Morgan and P. Burrill for their excellent work on our target. We
are grateful to H. Fearing for access to his RMC computer code and
helpful discussions. This experiment was supported by the NSERC and
NRC of Canada, the NSF of the USA, the PSI of Switzerland, and the ARC
of Australia.

\epsfysize=8cm
\begin{figure}[hbt]
\epsffile{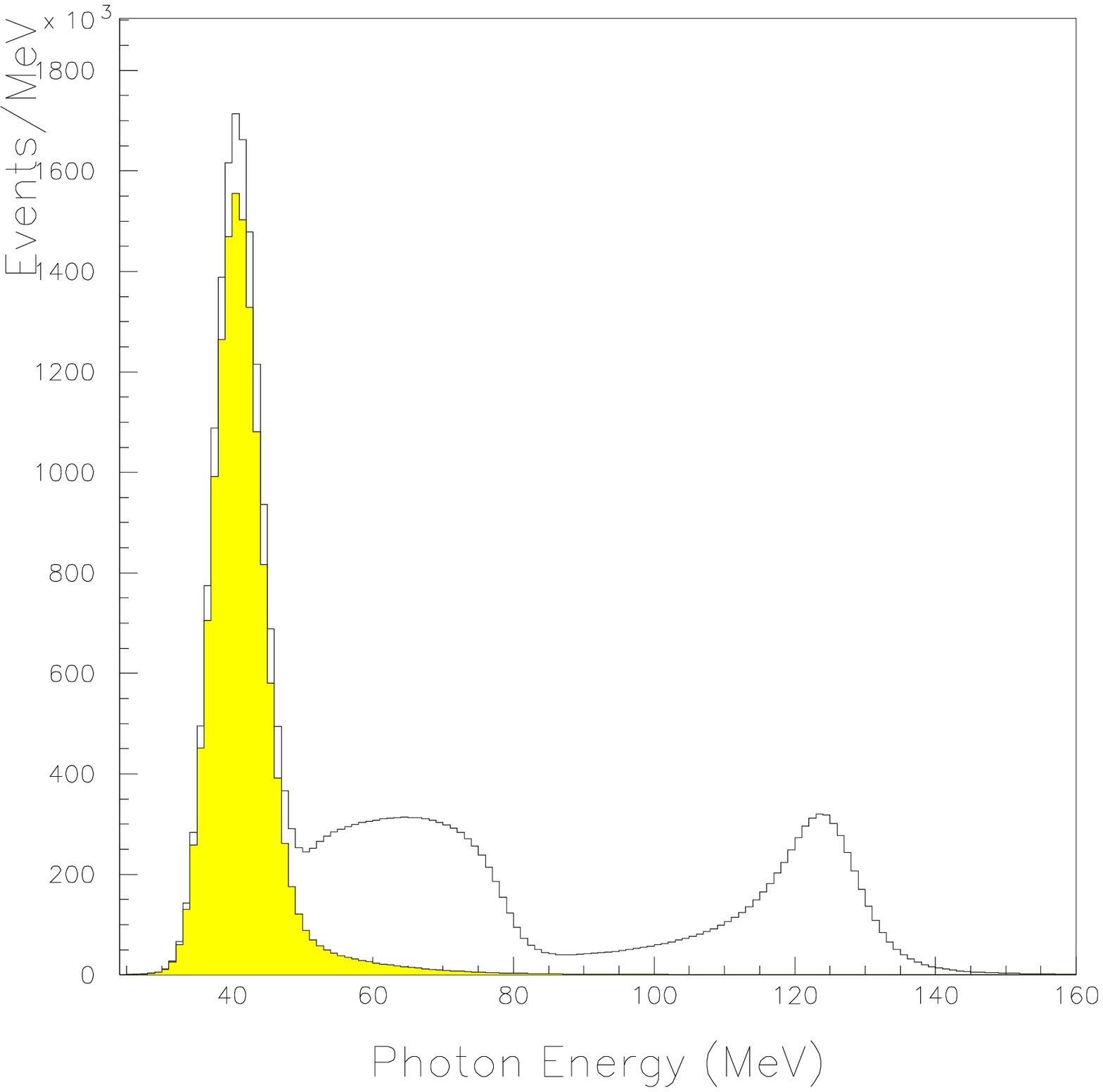}
\caption{Photon energy spectrum before cuts are applied; 
both $\mu^{-}$ and $\pi^{-}$ induced events
are observed. Shaded region: prompt cuts applied, pion events 
are eliminated.}
\end{figure}

\epsfysize=8cm
\begin{figure}[hbt]
\epsffile{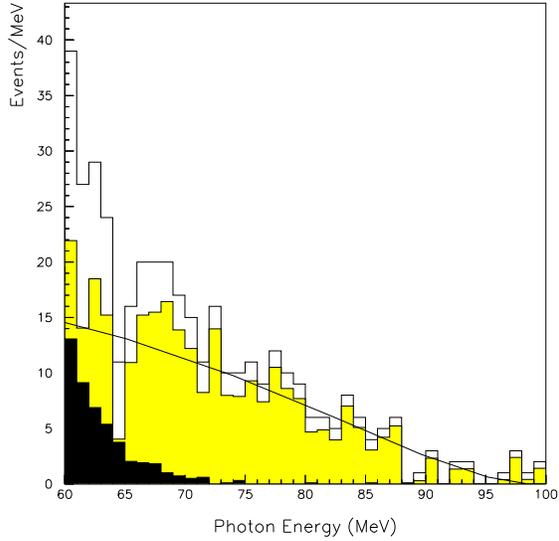}
\caption{Open histogram: Photon energy spectrum after all cuts were 
applied. Shaded histogram: final spectrum after background subtraction.  
Dark shaded overlay: normalized $\mu^{+}$ (annihilation-in-flight 
removed) bremsstrahlung background spectrum.
Curve: theoretical spectrum \protect\cite{Beder} 
for the best-fit
value of $g_p/g_a$.}
\end{figure}

\epsfysize=8cm
\begin{figure}[hbt]
\epsffile{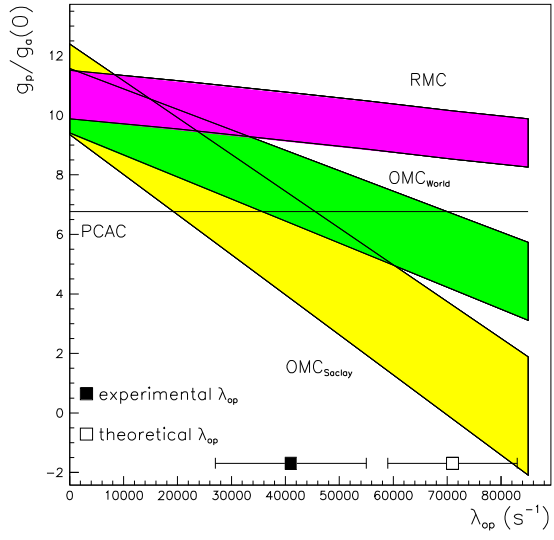}
\caption{$g_{p}(q^2)/g_a(0)$, evaluated at $q^{2} = -0.88m_{\mu}^{2}$, 
vs. the ortho to para transition rate $\lambda_{op}$ for OMC and RMC on 
hydrogen (see discussion in text).}
\end{figure}

\end{document}